\documentclass[11pt]{article}
\usepackage[utf8]{inputenc}
\usepackage{fullpage}
\usepackage{amsmath,amsfonts,amssymb,amsthm}
\usepackage{textcomp}
\usepackage{todonotes}
\usepackage{url}
\usepackage{enumitem}

\usepackage{thm-restate}

\usepackage{enumitem}

\mathchardef\mhyphen="2D

\usepackage{algorithm}
\usepackage{algpseudocode}
\algtext*{EndWhile}
\algtext*{EndIf}
\algtext*{EndFor}

 \newtheorem{theorem}{Theorem}[section]
 \newtheorem{lemma}[theorem]{Lemma}
 \newtheorem{claim}[theorem]{Claim}

\newcommand{\poly}{\mathrm{poly}}
 \newtheorem{definition}[theorem]{Definition}
 
\newcommand{\ceil}[1]{{\lceil#1\rceil}}

\graphicspath{{figs/}}

\newcommand{\opt}{\mathrm{opt}}
\newcommand{\cost}{\mathrm{cost}}

\renewcommand{\root}{\mathrm{root}}

\newcommand{\bfe}{\mathbf{e}}
\newcommand{\bfT}{\mathbf{T}}

\newcommand{\bfV}{\mathbf{V}}
\newcommand{\bfO}{{\mathbf{O}}}
\newcommand{\bfr}{{\mathbf{r}}}

\DeclareMathOperator*\E{\mathbb{E}}

\newcommand{\bfE}{\mathbf{E}}

\newcommand{\lbl}{\mathrm{label}}

\newcommand\genstatetree{\mathrm{gen\mhyphen state\mhyphen tree}}

\newcommand{\Z}{\mathbb{Z}}
\newcommand{\R}{\mathbb{R}}

\newcommand{\cnstrTzero}{\mathrm{cnstr\mhyphen}\bfT^\circ}

\newcommand{\tail}{{\mathrm{tail}}}
\newcommand{\second}{{\mathrm{second}}}
\newcommand{\third}{{\mathrm{third}}}
\newcommand{\round}{{\mathrm{round}}}

\newcommand{\virtual}{{\mathrm{virt}}}
\newcommand{\state}{{\mathrm{state}}}
\newcommand{\base}{{\mathrm{base}}}

\newcommand{\set}[1]{\left\{#1\right\}}
\newcommand{\floor}[1]{\left\lfloor#1\right\rfloor}

\begin{document}
\title{On Approximating Degree-Bounded Network Design Problems}
%
%
\author{Xiangyu Guo\\ Dept.\ of Comp.\ Sci.\ and Eng.\\University at Buffalo, USA\\xiangyug@buffalo.edu \and
Guy Kortsarz\\ Dept.\  of Comp.\ Sci.\\Rutgers University Camden, USA\\guyk@camden.rutgers.edu  \and
Bundit Laekhanukit\\ITCS, \\SUFE, China\\bundit@sufe.edu.cn \and
Shi Li\\ Dept.\ of Comp.\ Sci.\ and Eng.\\University at Buffalo, USA\\shil@buffalo.edu \and
Daniel Vaz\\Operations Research Group, \\TU Munich, Germany\\daniel.vaz@tum.de \and
Jiayi Xian\\ Dept.\ of Comp.\ Sci.\ and Eng.\\University at Buffalo, USA\\jxian@buffalo.edu}
\date{}
%
%
\maketitle              

\begin{abstract}
	
	Directed Steiner Tree (DST) is a central problem in combinatorial optimization and theoretical computer science: Given a directed graph $G=(V, E)$ with edge costs $c \in \R_{\geq 0}^E$, a root $r \in V$ and $k$ terminals $K\subseteq V$, we need to output the minimum-cost arborescence in $G$ that contains an $r$\textrightarrow $t$ path for every $t \in K$.  Recently, Grandoni, Laekhanukit and Li, and independently Ghuge and Nagarajan, gave quasi-polynomial time $O(\log^2k/\log \log k)$-approximation algorithms for the problem, which are tight under popular complexity assumptions. 
		
	In this paper, we consider the more general Degree-Bounded Directed Steiner Tree (DB-DST) problem, where we are additionally given a degree bound $d_v$ on each vertex $v \in V$, and we require that every vertex $v$ in the output tree has at most $d_v$ children.  We give a quasi-polynomial time $(O(\log n \log k), O(\log^2 n))$-bicriteria approximation: The algorithm produces a solution with cost at most $O(\log n\log k)$ times the cost of the optimum solution that violates the degree constraints by at most a factor of $O(\log^2n)$. This is the first non-trivial result for the problem.

	While our cost-guarantee is nearly optimal, the degree violation factor of $O(\log^2n)$ is an $O(\log n)$-factor away from the approximation lower bound of $\Omega(\log n)$ from the set-cover hardness. 
	The hardness result holds even on the special case of  the {\em Degree-Bounded Group Steiner Tree} problem on trees (DB-GST-T).
	With the hope of closing the gap, we study the question of whether the degree violation factor can be made tight for this special case. We answer the question in the affirmative by giving an $(O(\log n\log k),
	O(\log n))$-bicriteria approximation algorithm for DB-GST-T.  	
\end{abstract}
 \thispagestyle{empty}
\newpage

\setcounter{page}{1}

\section{Introduction} 
\label{sec:intro}

Network design is a central problem in combinatorial optimization and computer science. 
To capture more practical situations, 
the more general model of network design with \emph{degree-constraints} was suggested in the early 90's \cite{RaviMRRH93,FurerR94} and has attracted researchers in both theory and practice for decades. 
%
%
One of the most famous examples is the {\em Degree-Bounded Minimum Spanning Tree} (DB-MST)
problem, which models the problem of designing a multi-casting network in which each node only has enough power to broadcast to a bounded number of its neighbors. This problem has been studied in a sequence of works (see, e.g.,\cite{KonemannR02,KonemannR05,Goemans06,SinghL15}), leading to the breakthrough result of Goemans \cite{Goemans06} followed by the work of Singh and Lau \cite{SinghL15}, which settled down the problem by giving an algorithm that outputs a solution with optimum cost, while violating the degree bound by an additive factor of +1~\cite{SinghL15}. %
Since the works on DB-MST, many works have been dedicated to the study the generalizations of the problem: the {\em Degree-Bounded Steiner Tree} problem, in which the goal is to find a minimum-cost subgraph that connects all the terminals, while meeting the given degree bounds, was studied in  \cite{KonemannR03-Steiner,LauS13}. The {\em Survivable  Network Design} problem, where each pair of nodes $v,w$ are required to have at least $\lambda_{vw}$  edge-disjoint $v$-$w$ paths, has also been studied in literature; see, e.g., \cite{LauNSS09,LauS13}. %

Recently, degree-bounded network design problems have been studied in the online setting \cite{DehghaniEHL16,DehghaniEHLRS16,DehghaniEHLS18}. Besides the standard (also called point-to-point) network design problems, Dehghani~et~al.~\cite{DehghaniEHL16} also studied the {\em Degree-Bounded Group Steiner Tree} problem (DB-GST). They gave a negative result, which shows that it is not possible to approximate both cost and weight of the Online DB-GST problem simultaneously, even when the input graph is a star. More specifically, there exists an input demand sequence that forces any algorithm to pay a factor of $\Omega(n)$ either in the cost or in the degree violation. %
To date there was no non-trivial approximation algorithm for DB-GST, either in the online or offline setting, and even when all the edges have zero-cost. This was listed as an open problem by Hajiaghayi~\cite{Hajiaghayi-FND16} at the 8th Flexible Network Design Workshop (FND 2016).

In this paper, we study a degree-bounded variant of the classic network design problem, the {\em Degree-Bounded Directed Steiner Tree} problem (DB-DST). Formally, in DB-DST, we are given an $n$-vertex directed graph $G=(V,E)$ with costs on edges, a root vertex $r$, a set of $k$ terminals $K$, and degree bounds $d_v$ for each vertex $v$.
The goal is to find a minimum-cost rooted tree $T\subseteq G$ that contains a path from the root $r$ to every terminal $t\in K$, while respecting the degree bound, i.e., the out-degree of each vertex $v$ in $T$ is at most $d_v$. 
Despite being a classic problem, there was no previous positive result on DB-DST as it is a generalization of DB-GST.

The barriers in obtaining any non-trivial approximation algorithm for DB-GST and DB-DST are similar. %
Most of the previous algorithms to these two problems either run on the metric closure of the input graph~\cite{GargKR00,FriggstadKKLST14,Rothvoss11}, require metric-tree embedding~\cite{GargKR00,Bartal96,FakcharoenpholRT04} or use height-reduction techniques~\cite{Zelikovsky97,CharikarCCDGGL99,GrandoniLL19,GhugeN18}, all of which lose track of the degree of the solution subgraph.

We solve the open problem of Hajiaghayi~\cite{Hajiaghayi-FND16}, by presenting a bi-criteria $(O(\log k\log n),O(\log^2 n))$-approximation algorithm for DB-DST that runs in quasi-polynomial-time (see Section~\ref{sec:result} for the definition). 
Our technique expands upon the recent result of Grandoni, Laekhanukit and Li~\cite{GrandoniLL19} for the Directed Steiner Tree problem. 
We observe that the algorithm in \cite{GrandoniLL19} can be easily extended to the problem with degree bounds. 
Nevertheless, to amend the degree-constrained problem into their framework, we are required to prove a concentration bound for the degrees, which is rather non-trivial. Notice that the $O(\log n\log k)$-approximation factor on the cost of the tree is almost tight due to the hardness of $\Omega(\log^{2-\epsilon}n)$ in \cite{HalperinK03} for Directed Steiner Tree and the slightly improved hardness of $\Omega({\log^2n}/\log\log n)$ in \cite{GrandoniLL19}.


While our result for DB-DST is (almost) tight on the cost guarantee, the degree violation factor $O(\log^2n)$ is an $O(\log n)$ factor away from the approximation lower bound of $\Omega(\log n)$ from the set-cover hardness. 
To understand if the gap can be reduced, we study the special case of DB-DST obtained from the hardness construction in \cite{HalperinK03}, namely the {\em Degree-Bounded Group Steiner Tree problem on trees} (DB-GST-T). In this problem, we are given an (undirected) tree $T^\circ=(V^\circ,E^\circ)$ with edge-costs, a root $r$, $k$ subsets of vertices (called groups) $O_1,\ldots,O_k\subseteq V$ and a degree bound $d_v$ for each vertex $v\in V^{\circ}$. The goal is to find a minimum-cost subtree $T\subseteq T^\circ$ that joins $r$ to at least one vertex from each group $O_t$, for every $t\in[k]$, while respecting the degree bound, i.e., the number of children of each vertex $v$ in $T$ is at most $d_v$.
We present an $(O(\log k\log n), O(\log n))$-bicriteria approximation algorithm for DB-GST-T.  So, the degree violation of our algorithm is tight and the cost-guarantee is almost tight. 
This improves upon the $O(\log n\log k, \log n\log k)$-bicriteria approximation algorithm due to Kortsarz and Nutov \cite{KortsarzN2020} who observe that the randomized rounding algorithm in \cite{GargKR00} also gives a guarantee on degree-violation.

\subsection{Our Results} \label{sec:result}
Our first result is an $(O(\log k\log n), O(\log^2 n))$-bicriteria approximation for DB-DST that runs in quasi-polynomial time: 
%
We say that a randomized algorithm is an $(\alpha, \beta)$-bicriteria-approximation algorithm if it outputs a tree $T$ containing an $r$\textrightarrow $t$ path for every terminal $t \in K$ such that the number of children of every vertex $v$ in $T$ is at most $\beta\cdot d_v$, and the expected cost of the tree is at most $\alpha$ times the cost of the optimum tree that does not violate the degree constraints.  
\begin{theorem}
	\label{thm:main-DB-DST}
	There is a randomized $(O(\log  n \log k), O(\log^2 n))$-bicriteria approximation algorithm for the degree-bounded directed Steiner tree problem in $n^{O(\log n)}$-time. 
\end{theorem}

To the best of our knowledge, our result for DB-DST is the first non-trivial bicriteria approximation for the problem.  As we mentioned, the $O(\log n\log k)$-factor for the cost is almost tight due to the hardness results of \cite{HalperinK03} and \cite{GrandoniLL19} for DST.  There is a hardness of $\Omega(\log n)$ for the degree violation factor from the set-cover problem, even if we ignore the cost of the output tree. 

\paragraph{Remark} As in \cite{GrandoniLL19,GhugeN18}, we could save a factor of $\log \log n$  in the approximation factor for the problem, with a slight increase in the running time. However, this complicates the algorithmic framework. To deliver the algorithmic idea in a cleaner way, we choose to present the results with $O(\log n \log k)$ approximation ratios.   \medskip

Our second result is for the degree-bounded group Steiner tree problem on trees (DB-GST-T). We obtain an $\big(O(\log n\log k), O(\log n)\big)$-bicriteria approximation, which is (almost) tight on both factors:

\begin{restatable}{theorem}{thmdbgstt}
	\label{thm:dbgst-t}
	There is a randomized $\big(O(\log n\log k), O(\log n)\big)$-bicriteria approximation for the degree-bounded group Steiner tree problem on trees. 
\end{restatable}

\subsection{Our Techniques}
\label{sec:intro:techniques}

Our algorithm for degree-bounded directed Steiner tree takes ingredients from both \cite{GrandoniLL19} and \cite{GhugeN18}. 
As in these papers, we consider an optimum solution, and recursively partition it into balanced sub-trees; we then assign a ``state'' to each of these sub-trees.
The tree structure of this recursive partition, as well as all of the states, form what we call a \emph{state tree}.
We solve the problem indirectly, by finding a good state tree, which we can transform back into a corresponding good solution.
The state of a sub-tree contains a set of special vertices in the sub-tree that we call \emph{portals}; these were used in \cite{GhugeN18} to obtain their improved approximation algorithm for DST. 
We construct a super-tree $\bfT^\circ$ that contains all possible state trees as sub-trees and reduce the problem considered into that of finding a good sub-tree of small cost in $\bfT^\circ$.  This can be done by formulating a linear program (LP) relaxation and rounding the LP solution using a recursive procedure. The construction of the super-tree and the LP rounding techniques are similar to those in \cite{GrandoniLL19}.  To extend the algorithm to DB-DST, 
we need to store the degrees of all of the portals in the state.


This algorithmic framework outputs a so-called ``multi-tree'': This is a tree where a vertex or an edge can appear multiple times. Repeating the procedure for $Q = O(\log n \log k)$ times, we obtain a set of $Q$ multi-trees.
This process violates the degree requirements and thus we obtain bicriteria approximation results. The analysis of this process is non-trivial as we need to prove a concentration bound on the number of times a vertex appears in a multi-tree. 


Our technique for DB-GST-T is in observing that the rounding algorithm for GST-T (no degree bounds) in \cite{GargKR00} is indeed a generalization of random walk. As we slightly boost the branching probability by a constant factor, this (almost) does not affect the degree bound, but the probability of connecting the root vertex to each group is amplified dramatically. A drawback is that it also incurs a huge blow-up in the cost. 
To handle the blow-up, we stop amplifying the branching probability when the connecting probability is sufficiently large. 
The best (but inaccurate) way to illustrate our algorithm is by considering a random walk from the root vertex to a group $O_t$. We change the random process by branching into two directions simultaneously in each step, and then stop the extra branching when it generates $\Theta(\log n)$ simultaneous random walks. Since we have $O(\log n)$ simultaneous random walks, the cost incurred by the process is blown-up by a factor $O(\log n)$, but the degree-violation is blown-up by only a factor $2$. At the same time, the probability of reaching the group $O_t$ goes up by a factor $\Omega(\log n)$. Thus, if we need $O(\log k\log n)$ rounds to reach every group, then we now need only $O(\log k)$ rounds. There is no difference in the cost for running the algorithm for $O(\log k\log n)$ rounds or $O(\log k)$ rounds (with an extra $O(\log n)$ factor in the cost), but it saves a factor in the degree-violation of $O(\log n)$.




	\section{Preliminaries for Degree-Bounded Directed Steiner Tree} 
	\label{sec:prelim}

		 
	\subsection{Notations and Assumptions} 
	In our algorithm and analysis for the DB-DST problem, a tree is always an out-arborescence.  Given a tree $T$, we use $\root(T)$ to denote its root. Given $T$ and a vertex $v$ in $T$, we use $\Lambda_T(v)$ to denote the set of children of $v$, and $\Lambda^*_T(v)$ to denote the set of descendants of $v$ (including $v$ itself) in the tree $T$.  A sub-tree $T'$ of $T$ is a weakly-connected sub-graph of $T$; such a $T'$ must be an out-arborescence.  Sometimes, we shall use left and right children to refer to the two children of a vertex in a tree; in this case, the order of the two children is important and will be clearly specified.   For an edge $e=(u, v)$, we use $\tail(e) = v$ to denote its tail.  For a triple $\xi=(u, v, v')$ of three vertices, we use $\second(\xi) = v$ and $\third(\xi) = v'$ to denote the second and third parameter of $\xi$.

	Our input digraph is $G$. Let $d_{\max} = \max_{v \in V}d_v$. We shall assume each terminal $t \in K$ has only one incoming edge and no outgoing edges in $G$. This can be assumed w.l.o.g using the following simple operation: For every terminal $t \in K$ that does not satisfy the condition, we add a new vertex $t'$, an edge $(t, t')$ and replace $t$ with $t'$ in $K$. We increase $d_t$ by 1 and set $d_{t'} = 0$. 
	
	One more assumption we can make is that each non-terminal $u \in V \setminus K$ has at most  2 outgoing edges in $G$.  To make sure that this holds, we focus on some non-terminal $u$ with $b \geq 3$ outgoing edges. We replace the star centered at $u$ with its $b$ outgoing edges by a gadget which is a full binary-tree rooted at $u$ with $b$ leaves being the out-neighbors of $u$.  For every newly added vertex $u$, we set $d_u = d_{\max}$.  This way every vertex in $G$ will have at most $2$ outgoing edges. The cost of the edges in the gadget can be naturally defined. However, this operation changes the degree of vertices.  To address this issue, we define a simple transformation function $\phi_v: \Z \to \Z$ for every $v \in V$ as follows: If $v$ is a vertex in the original graph, then $\phi_v$ is identically 1. Otherwise, $v$ is a non-root internal vertex of some gadget and we define $\phi_v$ to be the identity function. Then we can compute the \emph{original degree} $\rho_u$ of a vertex $u$ in a tree $T$ of $G$ recursively as follows: $\rho_u = 0$ if $u$ is a leaf, and $\rho_u = \sum_{v \in \Lambda_T(u)}\phi_v(\rho_v)$ otherwise.   So, we require that for every $v$ in the output tree $T$, the original degree $\rho_v$ of $v$ is at most $d_v$. 
		
\subsection{Balanced Tree Partition}	
	We shall use the following basic tool as the starting point of our algorithm design. Its proof is elementary and deferred to Appendix~\ref{sec:appendix}. 
	\begin{lemma}
		\label{lemma:balanced-partition}
		Let $T = (V_T, E_T)$ be an $n$-vertex binary tree. 
		Then there exists a vertex $v\in V_T$ with $n/3 < |\Lambda^*_T(v)| \leq 2n/3+1$.  
	\end{lemma}

	Given a tree $T = (V_T, E_T)$ as  in the lemma, we can partition it into two trees $T_1 = (V_{T_1}, E_{T_1})$ and $T_2=(V_{T_2}, E_{T_2})$, where $T_2$ contains vertices in $\Lambda^*_T(v)$ and $T_1$ contains vertices in $V_T \setminus (\Lambda^*_T(v) \setminus \{v\})$.  First assume $n \geq 4$. Since $2n/3+1 < n$, we know that $v \neq \root(T)$, thus implying $\root(T_1) = \root(T) \neq \root(T_2) = v$, which is a leaf in $T_1$. Consequently, we have $E_{T_1}\uplus E_{T_2}=E_T$ and $V_{T_1} \cup V_{T_2} = V_T, V_{T_1}\cap V_{T_2}=\{\root(T_2)\}$. Moreover, $|V_{T_1}|, |V_{T_2}| \leq 2n/3+1$, which is strictly less than $n$. Thus, $T_1$ and $T_2$ are sub-trees that form a balanced partition of (the edges of) $T$.  We call this procedure the \emph{balanced tree partitioning} on $T$.
	
	When $n = 3$, there are 2 types of trees. If the root has two children, then we could not make both $|V_{T_1}|$ and $|V_{T_2}|$ to be smaller than $3$.  If the tree is a path of 2 edges, then we can choose $v$ to be the middle vertex and the procedure partitions the tree into two edges.  Later, we shall apply the balanced tree partitioning procedure recursively. We stop the recursion when the tree is either an edge, or only contains the root and its 2 children.  In other words, the tree has only 1 level of edges. 

	\subsection{Multi-Tree}
		We define a \emph{multi-tree} in $G$ as an intermediate structure. It is simply a tree over \emph{multi-sets} of vertices and edges in $G$:
		\begin{definition}[Multi-Tree]
			Given the input digraph $G = (V, E)$, a \emph{multi-tree} in $G$ is a tree $T = (V_T, E_T)$ where every vertex $a \in V_T$ is associated with a label $\lbl(a) \in V$ such that  for every $(a, b) \in E_T$, we have $(\lbl(a), \lbl(b)) \in E$.
		\end{definition}
		We say that each vertex $a \in V_T$ is a copy of the vertex $\lbl(a) \in V$ and each edge $(a, b) \in E_T$ is a copy of the edge $(\lbl(a), \lbl(b)) \in E$.  So, we say that $T$ is rooted at a copy of $v \in V$,  if $\lbl(\root(T)) = v$, and $T$ contains a copy of some $v \in V$ if there exists some $a \in V_T$ with $\lbl(a) = v$.  
		We extend the costs $c_e$, the functions $\phi_v$ and the degree bounds $d_v$  automatically to their copies in a multi-tree.  That means, for a vertex $a$ and an edge $(a,b)$ in a multi-tree, $d_a = d_{\lbl(a)}, \phi_a \equiv \phi_{\lbl(a)}$ and $c_{(a, b)}=c_{(\lbl(a), \lbl(b))}$. The cost of a multi-tree $T= (V_T, E_T)$ is naturally defined as $\cost(T) = \sum_{e \in E_T}c_{e}$.   Given a multi-tree $T$, the ``original degree'' $\rho_a$ of a vertex $a$ can be computed in the same way as before. 
		\begin{definition}[Good Multi-Trees]
				Let $T = (V_T, E_T)$ be a multi-tree in $G$.  We say that $T$ is good if it is rooted at a copy of $r$, has leaves being copies of terminals,  and the original degree of any vertex $a$ in $T$ is at most $d_a$.
		\end{definition}




We can then state the main theorem for DB-DST, which we prove in Sections \ref{sec:state}  to \ref{sec:round}.
\begin{theorem}[Main Theorem for DB-DST]
	\label{thm:main}
	There is an $n^{O(\log n)}$-time  randomized algorithm that outputs a good multi-tree $T = (V_T, E_T)$ such that 
	\begin{enumerate}[itemsep=0pt,label=(\ref{thm:main}\alph*),leftmargin=*]
		\item $\E_T[\cost(T)] \leq \opt$, where $\opt$ is the cost of the optimum solution for the instance.
		\item For every $t \in K$, we have $\Pr_T[V_T \text{ contains a copy of } t] \geq \Omega(1/\log n)$.
		\item For some $s = \Omega\left(\frac1{\log n}\right)$, it holds, for every $v \in V$, that
		\vspace*{-10pt}
		
		\begin{align*}
			\textstyle{\E\left[\exp\big(s\cdot(\text{number of copies of $v$ in $T$})\big)\right] \leq 1 + O\left(\frac1{\log n}\right)}.
		\end{align*} 
	\end{enumerate}	
\end{theorem}


We show that this implies Theorem~\ref{thm:main-DB-DST}. 
\begin{proof}[Proof of Theorem~\ref{thm:main-DB-DST}]
	%
	We run the algorithm in Theorem~\ref{thm:main} $Q$ times to obtain $Q$ good multi-trees $T_1, T_2, \cdots, T_Q$, for some large enough $Q =  O(\log n \log k)$.  Our output will contain all edges that appear in the $Q$ multi-trees. Notice that the output may not be a tree, but we can remove edges so that it becomes a tree. Applying union bound, all terminals appear in the union of the $Q$ trees with probability at least $0.9$, when $Q$ is big enough.  By Property~(\ref{thm:main}c) in the theorem statement, we have for every $v$,
	\begin{align*}
		\E\left[\exp\big(s\cdot(\text{\# copies of $v$ in $T_1, \cdots, T_Q$})\big)\right] &\leq \left(1+ O\left(\frac{1}{\log n}\right)\right)^{Q} = \exp(O(\log k)).
	\end{align*}
	The above inequality holds since the $Q$ trees are produced independently.  
	
	Thus, if $M = O(\log n)$ is big enough, by Markov's inequality we have 
	\begin{align*}
		\Pr\left[\exp\big(s\cdot(\text{\# copies of $v$ in $T_1, \cdots, T_Q$})\big) \geq \exp(M) \right] \leq \frac{1}{10n}.
	\end{align*}
	The event on the left side is exactly that the number of copies of $v$ in $T_1, \cdots, T_Q$ is at least $M/s$.  
	
	Thus, with probability at least $0.8$, every terminal $t$ appears in one of the $Q$ trees and every vertex $v$ appears at most $M/s = O(\log^2 n)$ times in $T_1, T_2, \cdots, T_Q$. Taking the union of all trees and reflecting the edges in original graph $G$, we have a sub-graph $G'$ of $G$ that contains a path from $r$ to every terminal $t \in K$. The total cost of edges in $G'$ is at most $O(\log n \log k)\cdot \opt$. For every vertex $v$, the out-degree of $v$ in $G'$ will be at most $(M/s)d_v = O(\log^2 n) d_v$.   We can take an arbitrary Steiner tree $T$ in $G'$ as the output of the algorithm. This gives us an $(O(\log n \log k), O(\log^2 n))$-bicriteria approximation algorithm for the degree-bounded directed Steiner tree problem. The running time of the algorithm is $n^{O(\log n)}$. 
\end{proof}
\vspace*{-15pt}

\paragraph{Organization} The remaining part of the paper is organized as follows. In Section~\ref{sec:state}, we define states and good state trees.  In Section~\ref{sec:reduction}, we argue that the problem of finding a small cost valid tree can be reduced to that of finding a small cost state-tree. In Section~\ref{sec:round}, we present our linear programming rounding algorithm that finishes the proof of Theorem~\ref{thm:main}.  
 Section~\ref{sec:dbgst-t} is dedicated to the proof of Theorem~\ref{thm:dbgst-t} for the degree-bounded group Steiner tree problem on trees (DB-GST-T).

\section{States and State-Trees}
\label{sec:state}
	Given the optimum tree $T^*$ (which is binary by our assumptions) for the DB-DST problem, we can apply the balanced tree partitioning recursively to obtain a \emph{decomposition tree}: We start from $T^*$ and partition it into two trees $T_1$ and $T_2$ using the balanced-tree-partitioning procedure, and then recursively partition $T_1$ and $T_2$ until we obtain sub-trees with 1 level of edges: Such a tree contains either a single edge, or  two edges from the root. Then the decomposition tree is a full binary tree where each node corresponds to a sub-tree of $T^*$.   Due to the balance condition, the height of the tree will be $O(\log n)$. Throughout the paper, we shall use $h = \Theta(\log n)$ to denote an upper bound on the height of this decomposition tree.
	
	Thanks to its small depth, the decomposition tree becomes the object of interest. However, as each node in the tree corresponds to a sub-tree of the optimum solution $T^*$, it contains too much information for the algorithm to handle. Instead, we shall only extract a small piece of information from each node that we call the \emph{state} of the node.  On one hand,  a state contains much less information than a sub-tree does, so we can afford to enumerate all possible states for a node. On the other hand, the states of nodes in the decomposition tree still contain enough information for us to check whether the correspondent multi-tree is good.  We call the binary tree of states a \emph{state tree}; we require in a \emph{good state tree}, the states of nodes satisfy some consistency constraints. 
	Then we can establish a two-direction connection between good multi-trees and good state trees. 
	
		Given a valid tree $T$ in $G$ and a sub-tree $T'$ of $T$, we now start to make definitions related to the state of $T'$ w.r.t $T$.  It is convenient to think that $T$ is the optimum tree $T^*$ and $T'$ is a sub-tree of $T = T^*$ obtained from the recursive balanced-partitioning procedure, since this is how we use the definitions.  However, the definitions are w.r.t general $T$ and $T'$; from now on till the end of Section~\ref{sec:state}, we fix any valid tree $T$ and its sub-tree $T'$.
	
	\subsection{Portals} 
		Other than $\root(T')$, the state for $T'$ w.r.t $T$ contains  the set of \emph{portals} of $T'$:
		\begin{definition}
			\label{def:portal}
			A vertex $v$ in $T'$ is a \emph{portal} in $T'$,  if $v$ is $\root(T')$ or a non-terminal leaf of $T'$.
		\end{definition}
		In general, the set of portals of $T'$ can be large, but if $T'$ is obtained from the recursive balanced-tree-partitioning procedure for $T$, then the number of portals can be shown to be at most $h+1$.  As we shall often use the root and set of portals together, we make the following definition:
	\begin{definition}[Root-Portals-Pair]
		$(r', S)$ is called a root-portals-pair if $r' \in S\subseteq V \setminus K$.
	\end{definition}
	
	It is easy to see that the root-portal-pairs for an internal node of the decomposition tree and its two children satisfy some properties stated in the following definition:
	\begin{definition}[Allowable Child-Pair] 
		\label{def:allowable-child-pair}
		Given three root-portals-pairs $(r', S), (r', S_1)$ and $(r'', S_2)$,  we say $((r', S_1), (r'', S_2))$ is an allowable child-pair of $(r', S)$ if $r'' \notin S, S_1 \cup S_2 = S \cup \{r''\}$ and $S_1 \cap S_2 = \{r''\}$.
	\end{definition}
	The following claim motivates the definition of allowable child pairs:
	\begin{claim}
		\label{claim:pair-is-allowable}
		Assume $T' =(V', E')$ contains at least 2 levels of edges. Let $T'_1 = (V'_1, E'_1)$ and $T'_2 = (V'_2, E'_2)$ be the two sub-trees obtained by applying the balanced tree partitioning  on $T'$.  Let $r' = \root(T') = \root(T'_1)$, $r'' = \root(T'_2) \neq r'$ and $S, S_1, S_2$ be the sets of portals in $T', T'_1, T'_2$ respectively. 
		Then,  $((r', S_1), (r'', S_2))$ is an allowable child-pair of $(r', S)$. 
	\end{claim}
	\begin{proof}
		First, $r''$ is not a portal of $T'$ since it is a non-root internal vertex in of $T'$.  Second, it is easy to see that $S_1 = (S \cup \{r''\}) \cap V'_1$ and $S_2 = (S \cup \{r''\}) \cap V'_2$. So, $S_1 \cup S_2 = S \cup \{r''\}$ and $S_1 \cap S_2 = \{r''\}$.
	\end{proof}
	
	
	\subsection{Degree Vectors}
	The next piece of the information in a state is a \emph{degree vector}:
	\begin{definition}
		 A degree vector for a set $S \subseteq V \setminus K$ is a vector $\rho = (\rho_v)_{v \in S}$, where $\rho_v$ is an integer in $[1, d_v]$ for every $v \in S$.
	\end{definition}
	Supposedly, $\rho_v$ will be the original degree of $v$ in the tree $T$.
	
	
	\begin{definition}[Consistency of degree vectors]
		\label{label:consistency}
		Given a root-portals-pair $(r', S)$, an allowable child-pair $((r', S_1), (r'', S_2))$ of $(r', S)$, three degree vectors $\rho, \rho^1$ and $\rho^2$ for $S,  S_1$ and $S_2$ respectively, we say $\rho^1$ and $\rho^2$ are consistent with $\rho$, if 
		\begin{itemize}[itemsep=0pt]
			\item for every $v \in S_1 \setminus \{r''\}$, we have $\rho_v = \rho^1_v$, 
			\item for every $v \in S_2 \setminus \{r''\}$, we have $\rho_v =\rho^2_v$ and
			\item $\rho^1_{r''} =\rho^2_{r''}$.
		\end{itemize}
	\end{definition}
	So, the degree vectors are consistent if there is no contradictory information among them. 
	
	\begin{definition}[Edge/Triple Agreeing with Degree Vector]
		Given a root-portals-pair $(r', S)$ with $|S| \leq 2$, a degree vector $\rho$ for $S$, and an edge $(r', v) \in E$ with $\{r', v\} \setminus K = S$, we say $(r', v)$ agrees with $\rho$ if $\rho_{r'} =  (\phi_{v}(\rho_v)\text{ or }1)$, where $(\phi_v(\rho_v)\text { or } 1)$ denotes $\phi_v(\rho_v)$ if $\rho_v$ is defined (i.e, if $v \in S$) and $1$ otherwise.
		
		Similarly, given a root-portals-pair $(r', S)$ with $|S| \leq 3$, a degree vector $\rho$ for $S$, and two edges $(r', v), (r', v') \in E$ such that $\{r', v, v'\} \setminus K = S$, we say the triple $(r', v, v')$ agrees with $\rho$ if $\rho_{r'} = (\phi_{v}(\rho_v)\text{ or }1)+ (\phi_{v'}(\rho_{v'})\text{ or }1)$.
	\end{definition}
	
	Notice that in the above definition either $v \in S$ or $v \in K$. In the former case, $\rho_v$ is defined; in the latter case $\rho_v$ is not defined but we know $\phi_v$ is identically 1. The same argument holds for $v'$.  The definition corresponds to the case when $T'$ is a base case of the recursive balanced tree partitioning, i.e., $T'$ contains only 1 level of edges.  If $T'$ contains an edge $e = (r', v)$, then the portal set of $T'$ is $\{r', v \} \setminus K$. We shall have $\rho_{r'} = \phi_{v}(\rho_v)\text{ or } 1$.  Thus, if $\rho$ is restricted to the portal set, we have $\rho_{r'} = (\phi_{v}(\rho_v)\text{ or }1)$. Similarly, if $T'$ contains 3 vertices $(r', v, v')$ with $r'$ being the root, then we must have $\rho_{r'} = (\phi_{v}(\rho_v)\text{ or }1)+ (\phi_{v'}(\rho_{v'})\text{ or }1)$.
	
	\subsection{States and Good State-Trees}
	With degree vectors, we can define states and good state-trees:
	\begin{definition}
	\label{def:state}
	A \emph{state} is a tuple $(r', S, \rho)$ where $(r', S)$ is a root-portals-pair and $\rho$ is a degree vector for $S$. 
	
	The state of the tree $T'$ w.r.t $T$ is the tuple $(r', S, \rho)$ with $r' = \root(T')$, $S$ being the set of portals in $T'$, and $\rho$ being the vector of original degrees of vertices in $S$ w.r.t the tree $T$.
	\end{definition}
	
	\begin{definition}[Good State Trees]
	\label{def:good-state tree}
	\emph{A good state tree} is a full binary tree $\tau$ of depth at most $h$, where every node ${p}$ is associated with a state $(r'_{p}, S_{p}, \rho^{p})$, and every leaf ${o}$ is associated with either an edge $e_{o} \in E$ or a triple $\xi_o$ such that the following conditions hold. 
	\begin{enumerate}[leftmargin=*,label=(\ref{def:good-state tree}\alph*),itemsep=0pt]
		\item $\left(r'_{\root(\tau)}, S_{\root(\tau)}\right) = (r, \{r\})$.  \label{property:gst-root-portal}
		\item For any leaf ${o}$ of $\tau$, either $e_{o}$ or $\xi_o$ agrees with $\rho^{o}$.  \label{property:gst-agree}
		\item For an internal node ${p}$ in $\tau$, letting ${q}$ and ${o}$ be the left and right children of ${p}$, then the pair $((r'_{q}, S_{q}), (r'_{o}, S_{o}))$ is an allowable child-pair of $(r'_{p}, S_{p})$ (so, $r'_{q} = r'_{p} \neq r'_o$), and $\rho^{q}$ and $\rho^{o}$ are consistent with $\rho^{p}$.  \label{property:gst-consistency}
	\end{enumerate}

	We say that a terminal $t \in K$ is \emph{involved} in a good state tree $\tau$ if there exists a leaf ${o}$ of $\tau$ with $t  =\tail(e_{o})$, or $t \in \{\second(\xi_o), \third(\xi_o)\}$.
	\end{definition}
	
	Given a good state tree $\tau$, and a leaf ${o}$ in $\tau$, we define the cost $c({o})$ as follows. If $e_o$ is defined, then we define $c(o) = c_{e_o}$; otherwise, define $c(o) = c_{(r'_o, \second(\xi_o))} + c_{(r'_o, \third(\xi_o))}$. 
	The cost of a state-tree $\tau$ is defined as $\cost(\tau):=\sum_{{o}\text{ leaf of }\tau}c({o})$.

\section{Reduction to Finding Good State-Trees}
\label{sec:reduction}
\subsection{From a Valid Tree to a Good State-Tree Involving All Terminals}


In this section, we show that the decomposition tree of the optimum tree $T^*$ can be turned into a good state tree $\tau^*$ with cost $\cost(\tau^*) = \cost(T^*)$ that involves all terminals. As we alluded, the state tree $\tau^*$ is constructed by taking the state for each node in the decomposition tree for $T^*$.  Formally, it  is obtained by calling $\genstatetree(T^*)$ (defined in Algorithm~\ref{alg:gen-state tree}). In the algorithm $\rho^{T^*}$ is the vector of original degrees of all vertices in $T^*$.
The procedure is only for analysis purpose; it is not a part of our algorithm. 
\begin{algorithm}[H]
		\caption{$\genstatetree(T')$}
		\label{alg:gen-state tree}
		\begin{algorithmic}[1]
			\State create a node ${p}$ with $r'_p = \root(T'), S_p =$ portals of $T'$ and $\rho^p$ being $\rho^{T^*}$ restricted to $S_p$
			\If{$T'$ has only 1 level of edges} 
				\State \textbf{if} $T'$ contains a single edge $e$ \textbf{then} let $e_{p} =e$ and \Return the single node ${p}$
				\State \textbf{otherwise}, $T'$ contains two edges $(r', v)$ and $(r', v')$, let $\xi_p = (r', v, v')$ and \Return $p$
			\EndIf
			\State apply balanced tree partitioning to decompose $T'$ into $T'_1$ and $T'_2$
			\State $\tau_1 \gets \genstatetree(T'_1), \tau_2 \gets \genstatetree(T'_2)$
			\State \Return the tree $\tau$ obtained by combining ${p}, \tau_1$ and $\tau_2$ with edges $({p}, \root(\tau_1))$ and $({p},\root(\tau_2))$, with $\root(\tau_1)$ and $\root(\tau_2)$ being the left and right children of ${p}$ respectively
		\end{algorithmic}
\end{algorithm}	

\begin{lemma}
	\label{lemma:valid-tree-to-state tree}
	$\tau^*$ is a good state tree involving  all terminals and $\cost(\tau^*) = \cost(T^*)$. 
\end{lemma}
\begin{proof}
	We first show that $\tau^*$ is a good state tree, by showing that it satisfies all the properties in Definition~\ref{def:good-state tree}. Property~\ref{property:gst-root-portal}  trivially holds by the way we define the parameters for the root recursion of $\genstatetree$.   Property~\ref{property:gst-agree} holds by that each $\rho^p$ is $\rho^{T^*}$ restricted to $S^p$.  Property~\ref{property:gst-consistency} follows from the same facts and Claim~\ref{claim:pair-is-allowable}. 
	$\cost(\tau^*) =  \sum_{e \in E_{T^*}}c_e = \cost(T^*)$ since every edge in $T^*$ counted exactly once in $\tau^*$.
\end{proof}

\subsection{From a Good State Tree to  a Good Multi-Tree} \label{subsec:state-tree-to-multi-tree}
Now we focus on the other direction of the reduction. Suppose we are given a good state tree $\tau$, and our goal is to construct a good multi-tree $T$ with $\cost(T) = \cost(\tau)$. Moreover, if a terminal $t \in K$ is involved in $\tau$, then $T$ contains a copy of $t$.

	The multi-tree $T$ is constructed by joining the edges associated with all leaf nodes ${o}$ in $\tau$ using a recursive procedure.  For each node ${p}$ in $\tau$ we shall construct a multi-tree $T_{p}$ for ${p}$, as well as a mapping $\pi_{p}$ from $S_{p}$ to vertices in $T_{p}$.  The multi-tree $T_{p}$ and the mapping $\pi_{p}$ satisfy the following properties:
\begin{enumerate}[leftmargin=*, label=(P\arabic*),itemsep=0pt]
	\item For every $v \in S^{p}$, we have $\lbl(\pi_{{p}}(v)) = v$; that is, $\pi_{{p}}(v)$ is a copy of $v$.
	\item $\pi_{p}(r'_{p}) = \root(T_{p})$.
\end{enumerate}
In particular, the two properties imply that $\root(T_{p})$ is a copy of $r'_{p}$. 

The trees and mappings are constructed from the bottom to the top of the tree $\tau$. Focus on a leaf node ${p}$ with $e_{p} = (r', v)$. If $e_p$ is defined, then $T_{p}$ only contains a copy of the edge $(r', v)$. $\pi_{p}$ maps $r'$ to the copy of $r'$, and if $v \notin K$ (thus, $v \in S_p$), $v$ to the copy of $v$ in $T_{p}$.  Otherwise $\xi_p$ is defined.  Then $T_{p}$ contains a tree with two edges: a copy of $(r'_p, \second(\xi_p))$ and a copy of $(r'_p, \third(\xi_p))$. $\pi_p$ can also be defined naturally. 

Now consider the case that ${p}$ is an internal node and let ${q}$ and ${o}$ be its left and right children.  Then, we have $r'_{p} = r'_{q}, r'_o \notin S_p, S_{q} \cup S_{o} = S_{p} \cup \{r'_{o}\}$ and $S_{q} \cap S_{o} = \{r'_{o}\}$ by Property~\ref{property:gst-consistency}.  Then we identify $\pi_{{q}}(r'_{o})$ with $\pi_{{o}}(r'_{o}) = \root(T_{{o}})$, and then the multi-tree $T_{p}$ is the new tree containing vertices in $T_{{q}}$ and $T_{{o}}$. Notice that both $\pi_{{q}}(r'_{o})$ and $\pi_{{o}}(r'_{o})$ are copies of $r'_{o}$; thus the obtained $T_{p}$ can be well-defined.  The mapping $\pi_{p}$ is just the combination of $\pi_{{q}}$ and $\pi_{{o}}$: For a vertex $v \in S_{q}$, let $\pi_{p}(v) = \pi_{{q}}(v)$; for a vertex $v \in S_{o}$, let $\pi_{p}(v) = \pi_{{o}}(v)$; since $S_{q} \cap S_{o} = \{r'_{o}\}$ and we identified $\pi_{{q}}(r'_{o})$ with $\pi_{{o}}(r'_{o})$, the mapping is well-defined. Also, it is easy to see that (P1) and (P2) holds for $T_{p}$  and $\pi_{p}$.
	 
Our final multi-tree for $\tau$ will be $T = T_{\root(\tau)}$.  It is straightforward to see that if $t \in K$ is involved in $\tau$, then $T$ contains a copy of $t$. 
Notice that all the $\rho^p$-vectors  are consistent with each other, and for every leaf $o$, $e_o$ or $\epsilon_o$ agrees with $\rho^o$. Thus, aggregating all the $\rho^p$ vectors will recover the vector $\rho^T$ of original degrees of vertices in $\rho^T$.  So, the multi-tree $T$ is good since every $v$ in $T$ has $\rho^T_v \in [1, d_v]$. The cost of $T$ is $\sum_{e \in E_T}c_e = \sum_{o:\text{ leaves of }\tau}c(o) = \cost(\tau)$.
\section{Finding a Good State Tree using LP Rounding}
\label{sec:round}
\subsection{Extended State Trees and Construction of $\bfT^0$}
With the relationship between good multi-trees and good state trees established, we can now focus on the problem of finding a good state-tree of small cost involving many terminals.  We shall construct a quasi-polynomial sized tree $\bfT^\circ$ so that every good state-tree $\tau$ corresponds a sub-tree $\bfT$ of $\bfT^\circ$ satisfying some property.  Roughly speaking, $\bfT^\circ$ is the ``super-set'' of all potential good state-trees $\tau$.  However, since the consistency conditions are defined over three states for a parent and its two children, it is more convenient to insert a ``virtual'' node between every internal node and its two children. Also, it is convenient to break a leaf state node $o$ into two nodes, one containing the state information and the other containing $e_o$ or $\xi_o$.  Formally, for a good state-tree $\tau$, we construct a correspondent tree $\bfT$ as follows.  
\begin{enumerate}[itemsep=0pt]
	\item Let $\bfT$ be a copy of $\tau$.  All nodes in $\bfT$ are called \emph{state nodes}. 
	\item For every internal state node $p$ in $\bfT$ with left and right children $p_1$ and $p_2$, we create a \emph{virtual node} $q$ and replace the two edges $(p, p_1)$ and $(p, p_2)$ with 3 edges $(p, q), (q, p_1)$ and $(q, p_2)$; $p_1$ is still the left child and $p_2$ is the right child.
	\item For every leaf state node $p$, we create a \emph{base node} $o$ and let $o$ be the child of $p$.  Then we move the $e_p$ or $\xi_p$ information from the node $p$ to node $o$: If $e_p$ is defined, then we let $e_o = e_p$ and undefine $e_p$; otherwise, let $\xi_o = \xi_p$ and undefine $\xi_p$. 
	\item We add a \emph{super node} $\bfr$ and an edge from $\bfr$ to the root of $\bfT$. $\bfr$ will be the new root for $\bfT$.
\end{enumerate}

 We call this $\bfT$ the \emph{extended state-tree} for $\tau$; we say $\bfT$ is good if its correspondent $\tau$ is good.  Clearly, there is a 1-to-1 correspondence between good state trees and good extended state trees.


Our $\bfT^\circ$ will be the ``super-set'' of all potential good extended state trees $\bfT$.   Formally, we create a super node $\bfr$ to be the root of $\bfT^\circ$. Then, for every $\rho_r \in [1, d_r]$, we call $\cnstrTzero(0, r, \{r\}, \rho = (\rho_r))$ to obtain a tree and let its root be a child of $\bfr$. 
	\begin{algorithm}[H]
		\caption{$\cnstrTzero(h', r', S, \rho)$}
		\begin{algorithmic}[1]
			\State create a state node $p$ with $(r'_p, S_p, \rho^p) = (r', S, \rho)$
			\For{every $(r', v) \in E$ such that $\{r', v\} \setminus K = S$ and $(r', v)$ agrees with $\rho$}
				\State create a ``base node'' $o$ with $e_o = (r', v)$ and let  $o$ be a child of $p$
				\State let $c(o) = c_{(r', v)}$ 
			\EndFor
			\For{every $(r', v), (r', v') \in E$ such that $\{r', v, v'\} \setminus K = S$ and $(r', v, v')$ agrees with $\rho$}
				\State create a ``base node'' $o$ with $\xi_o = (r', v, v')$ and let  $o$ be a child of $p$
				\State let $c(o) = c_{(r', v)}  + c_{(r', v')}$ 
			\EndFor			
			\If{$h' < h$}
				\For{every allowable child-pair $((r', S_1), (r'', S_2))$ of $(r', S)$}
					\For{every pair of degree vectors $\rho^1$ for $S_1$ and $\rho^2$ for $S_2$ such that $ \rho^1$ and $\rho^2$ are consistent with $\rho$}
						\State create a  ``virtual node''  $q$ and let $q$ be a child of $p$
						\State $\bfT_1 \gets \cnstrTzero(h'+1, r', S_1, \rho^1)$
						\State  $\bfT_2 \gets \cnstrTzero(h'+1, r'', S_2, \rho^2)$
						\State let the left and right sub-trees of $q$ be  $\bfT_1$ and $\bfT_2$ respectively 
					\EndFor
				\EndFor
			\EndIf
			\State \Return the tree $\bfT$ rooted at $p$
		\end{algorithmic}
	\end{algorithm}
	
	
	The following claim is immediate from the construction of $\bfT^\circ$.
	\begin{claim}
		\label{claim:good-extended-state-tree}
		A subtree $\bfT$ of $\bfT^\circ$ with $\root(\bfT) = \root(\bfT^\circ)$ is a good extended state tree if and only if the following happens:
		\begin{itemize}[itemsep=0pt]
			\item The super node in $\bfT$ has exactly one child (which is a state node).
			\item Each state node in $\bfT$ has exactly one child (which is an base node or a virtual node).
			\item For each virtual node $q$ in $\bfT$, both $q$'s children in $\bfT^\circ$ are in $\bfT$. 
		\end{itemize}
		On the other hand, every good extended tree $\bfT$ of depth at most $h+1$ is a sub-tree of $\bfT^\circ$ with root being $\root(\bfT^\circ)$.
	\end{claim}


	Also, we say that a vertex $v$ is involved in $\bfT$ if there is an base node $o$ in $\bfT$ with $v = \tail(e_o)$ or $v \in \{\second(\xi_o), \third(\xi_o)\}$. The cost of $\bfT$, denoted as $\cost(\bfT)$, is defined the sum of $c(o)$ over all base nodes in $\bfT$.  So, the problem now becomes finding a small-cost good extended state tree in $\bfT^\circ$ that involves each terminal with large probability. 

\subsection{LP Formulation}
	We formulate an LP relaxation for our task. Let $\bfV^\circ$ be the set of nodes in $\bfT^\circ$, $\bfr = \root(\bfT^\circ)$ and let $\bfV^\circ_\state, \bfV^\circ_\virtual$ and $\bfV^\circ_\base$ be the sets of state, virtual and base nodes in $\bfT^\circ$ respectively.  Notice that there is only one super node, which is the root $\bfr$.  For every $t \in K$, let $\bfO_t = \left\{t \in \bfV^\circ_\base: t=\tail(e_o)\text{ or } t\in \{\second(\xi_o), \third(\xi_o)\} \right\}$ be the set of base nodes involving $t$. Let $\bfT^*$ be our target good extended state tree; this is the tree correspondent to the good state tree $\tau^*$.  Then, in our LP,  we have a variable $x_p$ for every $p \in \bfV^\circ$, that indicates  whether $p$ is in the $\bfT^*$ or not. 
	\begin{equation}
		\min \qquad \sum_{o \in \bfV^\circ_\base}x_oc(o) \label{LP:DST}
	\end{equation}\vspace*{-20pt}
	
\noindent\begin{minipage}{0.55\textwidth}
	\begin{alignat}{2}
		\sum_{q \in \Lambda_{\bfT^\circ}(p)}x_q &= x_p, &\ &\forall p \in \bfV^\circ_\state \cup \{\bfr\} \label{LPC:DST-state} \\
		x_{p} &= x_q, &\  &\forall q \in \bfV^\circ_\virtual, p \in \Lambda_{\bfT^\circ}(q)\label{LPC:DST-virtual}\\[5pt]
		x_p &\in [0, 1], &\ &\forall p \in \bfV^\circ \label{LPC:DST-0-1}
	\end{alignat}
\end{minipage}
\begin{minipage}{0.44\textwidth}
	\begin{alignat}{2}
		\sum_{o \in \Lambda^*_{\bfT^\circ}(p) \cap \bfO_t} x_o &\leq x_p, &\  &\forall p \in \bfV^\circ,  t\in K  \label{LPC:DST-sum-o-for-t} \\
		\sum_{o \in \bfO_t } x_o &= 1, &\  &\forall t \in K \label{LPC:DST-t-covered} 
	\end{alignat}	
\end{minipage} \vspace*{10pt}

The objective function of LP~\eqref{LP:DST} is to minimize the total cost of all leaves in $\bfT^*$.  \eqref{LPC:DST-state} requires that for every state or super node $p$ in $\bfT^*$, exactly one child of $p$ is in $\bfT^*$. \eqref{LPC:DST-virtual} requires that a virtual node $q$ in $\bfT^*$ has both its children in $\bfT^*$.  \eqref{LPC:DST-sum-o-for-t} says for every node $p$ in $\bfT^*$ and every terminal $t \in K$, there is a most one descendant base node $o$ of $p$ that is in $\bfO_t$. In the whole tree $\bfT^*$, exactly one leaf node $o$ has $t = \tail(e_o)$ or $t \in \{\second(\xi_o), \third(\xi_o)\}$,  for every $t \in K$ (Constraint~\eqref{LPC:DST-t-covered}); in the LP, all the variables are between $0$ and $1$ (Constraint~\eqref{LPC:DST-0-1}).

Notice that \eqref{LPC:DST-sum-o-for-t} for $p = \bfr$ and any $t \in K$ and \eqref{LPC:DST-t-covered} for the same $t$ imply that $x_\bfr = 1$.  \eqref{LPC:DST-state} and \eqref{LPC:DST-virtual} imply that the $x$ values over the nodes of a root-to-leaf path in $\bfT^\circ$ are non-increasing.

\subsection{Rounding Algorithm}
Given a valid solution $x$ to LP~\eqref{LP:DST}, our rounding algorithm will round it to obtain set $\bfV \subseteq \bfV^\circ$, which induces a good state tree.   The algorithm is very similar to that of \cite{GargKR00} with the only one difference: For every state node or super-node $p$ that is added to $\bfV$, we add exactly one child $q$ of $p$ to $\bfV$, while the algorithm of \cite{GargKR00} makes independent decisions for each child. The algorithm is formally described in Algorithm~\ref{alg:GKR}.  In the main algorithm, we simply call $\round(\bfr)$.
	\begin{algorithm}
		\caption{$\round(p)$} \label{alg:GKR}
		\begin{algorithmic}[1]
			\If{$p \in \bfV^\circ_\state \cup \{\bfr\}$}
			\State randomly choose a child $q$ of $p$ according to probability vector $\left(\frac{x_q}{x_p}\right)_{q \in \Lambda_{\bfT^\circ}(p) }$
			\State \Return $\{p\} \cup \round(q)$
			\ElsIf{$p \in \bfV^\circ_\virtual$}
			\State \Return $\{p\} \cup \round(\text{left child of }p) \cup \round(\text{right child of }p)$		
			\Else
			\State \Return $\{p\}$
			\EndIf
		\end{algorithmic}
	\end{algorithm}
	It is straightforward to see that the tree induced by $\round(\bfr)$ is a good extended state tree.  The following claim also holds:
	\begin{claim}
		Let $p\in \bfV^\circ$ and $q \in \Lambda^*_{\bfT^\circ}(p)$. Let $\bfV$ be the random set returned by $\round(p)$. Then we have $\Pr[q \in \bfV] = \frac{x_q}{x_p}$.
	\end{claim}
	Applying the above claim for $p = \bfr$ and every $q \in \bfV^\circ_\base$, we have that the expected cost of the tree induced by $\bfV$ is exactly $\cost(x)$. 
		
	The main theorem we need about the rounding algorithm is as follows:
	\begin{theorem}
		\label{thm:terminal-probability}
		Let $\bfV$ be the random set returned by $\round(\bfr)$.  Then,  for any terminal $t \in K$ we have 
		\begin{align*}
			\Pr[\bfV \cap \bfO'_t \neq \emptyset] \geq \frac{1}{h+1}.
		\end{align*}
	\end{theorem}

	Theorem~\ref{thm:terminal-probability} was proved \cite{GargKR00} for the original rounding algorithm and was reproved in \cite{Rothvoss11}.  However, adapting the analysis to our slightly different rounding algorithm is straightforward and thus we omit the proof of the theorem here. 
	
	We now wrap up and finish the proof of the main theorem (Theorem~\ref{thm:main}) except for Property~(\ref{thm:main}c), which will be proved in Section~\ref{subsec:concentration}.
	
	  We solve LP\eqref{LP:DST} to obtain a solution $x$. Notice that $\cost(x) \leq \cost(\bfT^*) = \cost(\tau^*) = \cost(T^*)$.   
	Let $\bfV \gets \round(\bfr)$. Then by Claim~\ref{claim:good-extended-state-tree} and the rounding algorithm, the tree $\bfT$ induced by $\bfV$ is a good extended state tree.   Let $\tau$ be the good state tree correspondent to $\bfT$, and let $T$ be the good multi-tree in $G$ constructed using the procedure in Section~\ref{subsec:state-tree-to-multi-tree}.  The cost of the multi-tree $T$ is at most $\cost(x)$. By Theorem~\ref{thm:terminal-probability},  for every $t \in K$, the probability that $t$ is involved  $T$ is at least $1/(h+1) = \Omega(1/\log n)$. 
	
	Let us consider the running time of the algorithmic framework, which is polynomial on the size of the tree $\bfT^\circ$.  First notice that if $((r', S_1), (r'', S_2))$ is an allowable child pair of $(r', S)$, then we have $|S_1|, |S_2| \leq |S| + 1$ since $S_1 \cup S_2 = S \cup \{r''\}$.  Thus, a state-node $p$ at the $h'$-th level in $\bfT^\circ$ (the children of $\bfr$ have  level $0$ and for simplicity we do not consider super and virtual nodes when counting levels) has $|S_p|\leq h'+1$. Thus, every state node $p$ in $\bfT^\circ$ has $|S_p| \leq h + 1$. 
	
	Then we consider the degree of the tree $\bfT^\circ$, which is the maximum number of possible children of a state node $p$ with $(r'_p, S_p, \rho^p) = (r', S, \rho)$.  First, there are at most $n\times 2^{|S_p|} \leq n\cdot 2^{h+1}$ different allowable child pairs $((r', S_1), (r'', S_2))$ of the pair $(r', S)$: there are at most $n$ choices for $r''$ and $2^h$ ways to split $S$ into $S_1$ and $S_2$.  Then, for a fixed allowable child pair $((r', S_1), (r'', S_2))$ we consider the number of pairs of degree vectors $\big(\rho^1, \rho^2\big)$ such that $\rho^1$ and $\rho^2$ are consistent with $\rho$. This is determined by the value of $\rho^1_{r''} = \rho^2_{r''}$, which has at most $d_{\max}$ possibilities.  So, the number of virtual children of a state node is at most $n\cdot2^{h+1} \cdot d_{\max} = O(\poly(n)$ since $h = O(\log n)$. The number of child base nodes of $p$ is at most $n^2$.  Since the height of the tree $\bfT^\circ$ is at most $O(\log n)$, its size bounded by $(\poly(n))^{O(\log n)} = n^{O(\log n)}$. So the running time of the LP rounding algorithm is $n^{O(\log n)}$. This finishes the proof of Theorems~\ref{thm:main} except for Property~(\ref{thm:main}c).

	\subsection{Concentration Bound on Number of Copies of a Vertex Appearing in $T$}
	\label{subsec:concentration}
	Finally, we prove Property~(\ref{thm:main}c) in Theorem~\ref{thm:main}.  To this end, we shall fix a vertex $v \in V$.   For every vertex $p \in \bfV^\circ$, let $z_p = \sum_{o \in \Lambda^*_{\bfT^\circ}(p) \cap \bfO_v} x_o$. By Constraint~\eqref{LPC:DST-sum-o-for-t}, we have $z_p \leq x_p$.  Let $m_p = |\Lambda^*_{\bfT^\circ}(p) \cap \bfO_v \cap \bfV|$ be the total number of nodes in $\Lambda^*_{\bfT^\circ}(p) \cap \bfO_v$ that are selected by the rounding algorithm. 
	
	As is typical, we shall introduce a parameter $s > 0$ and consider the expectation the random exponential variables $\bfe^{s m_p}$ (we use $\bfe$ for the natural constant).  We shall bound $\E[\bfe^{sm_p}|p \in \bfV]$ from bottom to top by induction. So, in this proof, it is more convenient to for us to use a different definition of levels: the level of a node $p$ in $\bfT^\circ$ is the maximum number of edges in a path in $\bfT^\circ$ starting from $p$.  So, the leaves have level $0$ and for an internal node $p$ in $\bfT^\circ$, the level of $p$ is 1 plus the maximum of the level of $q$ over all children $q$ of $p$.  We define an $\alpha_i$ for every integer $i \geq 0$ as 
	$\alpha_0 = \bfe^s$ and $\alpha^i = \bfe^{\alpha_{i-1}-1}, \forall i \geq 1$.
	Notice that $\alpha_0, \alpha_1, \cdots$ is an increasing sequence.
	%
	Thus, we can induce the following lemma.
	\begin{lemma}
		\label{lemma:bound-exp-mp}
		For any node $p$ be in $\bfT^\circ$ of level at most $i$,
		$
			\E\Big[\bfe^{s m_p} \big| p \in \bfV \Big] \leq \alpha_i^{z_p/x_p}.
		$
	\end{lemma}
	\begin{proof}
			We prove the lemma by induction on $i$. If $i = 0$, then $p$ is a leaf, and thus, we have either $z_p = 0$ or $z_p = x_p$, depending on whether $p \in \bfO_v$ or not.  If $z_p = 0$, then $m_p$ is always $0$, and thus, $\E\Big[\bfe^{s m_p} \big| p \in \bfV \Big] = 1 = \alpha_0^{z_p/x_p}$.  If $z_p = x_p$, then $m_p$ is always $1$ (conditioned on $p \in \bfV$), and thus, $\E\Big[\bfe^{s m_p} \big| p \in \bfV \Big] = \bfe^s = \alpha_0^{z_p/x_p}$. So, the lemma holds if $i = 0$.
			
			Now, let $i \geq 1$ be any integer and we assume the lemma holds for $i-1$. We shall prove that it also holds for $i$.  Focus on a node $p$ of level at most $i$. Then all children $q$ of $p$ have level at most $i-1$. If $p$ is a virtual node, then $p \in \bfV$ implies that both children of $p$ in $\bfV$. Since the two children are handled independently in the rounding algorithm, we have 
			\begin{align*}
				\E\Big[\bfe^{sm_p}\big|p \in \bfV\Big]  &= \prod_{q \in \Lambda_{\bfT^\circ}(p)}\E \Big[\bfe^{sm_q} \big| p \in \bfV \Big] =\prod_{q \in \Lambda_{\bfT^\circ}(p)}\left[\frac{x_q}{x_p}\cdot \E[\bfe^{sm_q}|q \in \bfV] + 1-\frac{x_q}{x_p}  \right]\\
				&=\prod_{q \in \Lambda_{\bfT^\circ}(p)}\left[1 + \frac{x_q}{x_p} \Big(\E[\bfe^{sm_q}|q \in \bfV] -1\Big)  \right].
			\end{align*}
			If $p$ is the super node or a state node, then we have $\sum_{q \in \Lambda_{\bfT^\circ}(p)}x_q = x_p$. Conditioned on $p \in \bfV$, the rounding procedure adds exactly one child $q$ of $p$ to $\bfV$.  Then, we have 
			\begin{align*}
				\E\Big[\bfe^{sm_p}\big|p \in \bfV\Big]  &= \sum_{q \in \Lambda_{\bfT^\circ}(p)}\frac{x_q}{x_p} \E\Big[\bfe^{sm_q}\big|q \in \bfV\Big] 
				= 1 + \sum_{q \in \Lambda_{\bfT^\circ}(p)}\frac{x_q}{x_p}\Big( \E[\bfe^{sm_q}\big|q \in \bfV]-1\Big) \\
				&\leq \prod_{q \in \Lambda_{\bfT^\circ}(p)}\left[1 + \frac{x_q}{x_p} \Big(\E[\bfe^{sm_q}|q \in \bfV] -1\Big)  \right].
			\end{align*}
			Thus, we always have 
			\begin{align*}
				&\quad \E\Big[\bfe^{sm_p}\big|p \in \bfV\Big] 
				\leq \prod_{q \in \Lambda_{\bfT^\circ}(p)}\left[1 + \frac{x_q}{x_p} \Big(\E[\bfe^{sm_q}|q \in \bfV] -1\Big)  \right]\\
				&\leq \prod_{q \in \Lambda_{\bfT^\circ}(p)}\left[1+ \frac{x_q}{x_p}\big(\alpha_{i-1}^{z_q/x_q}-1\big)\right] &\text{by induction hypothesis} \\
				&\leq\exp\left[\sum_{q \in \Lambda_{\bfT^\circ}(p)}\frac{x_q}{x_p}\big(\alpha_{i-1}^{z_q/x_q}-1\big)\right] 
				\leq \exp\left[\frac{z_p}{x_p}(\alpha_{i-1}-1)\right] = \alpha_{i}^{z_p/x_p}. & \text{since $1 + \theta \leq e^\theta$ for every $\theta$}
			\end{align*}
	
	To see the second inequality in the last line, we notice the following three facts: (i) $\alpha_{i-1}^\theta - 1$ is a convex function of $\theta$ and when $\theta = 0$ its value is $0$, (ii) $z_q/x_q \in [0, 1]$ for every $q$ in the summation, and (iii) $\sum_{q\in \Lambda_{\bfT^\circ}(p)}\frac{x_q}{x_p}\cdot\frac{z_q}{x_q} = \frac{z_p}{x_p}$. So, the quantity inside $\exp(\cdot)$ has maximum value $\frac{z_p}{x_p}(\alpha_{i-1}^1-1)$. The equality in the last line is by the definition of $\alpha_i$. 
	\end{proof}
	
	Let $h' = \Theta(h) = \Theta(\log n)$ be the level of the root.  Now, we set $s = \ln(1+\frac{1}{2h'})$. We prove inductively the following lemma:
	\begin{lemma}
		\label{lemma:bound-alpha}
		For every $i \in[0, h']$, we have $\alpha_i \leq 1 + \frac{1}{2h' - i}$.
	\end{lemma}
	\begin{proof}
		By definition, $\alpha_0 = \bfe^s = 1+ \frac{1}{2h'}$ and thus the statement holds for $i = 0$.  Let $i \in [1, h']$ and assume the statement holds for $i-1$. Then, we have 
		\begin{align*}
		\alpha_i &= \bfe^{\alpha_{i-1}-1} \leq \bfe^{1 + \frac{1}{2h'-i+1}} \leq 1 + \frac{1}{2h'-i+1} + \left(\frac{1}{2h'-i+1}\right)^2\\
		&= 1 + \frac{2h'-i+2}{(2h'-i+1)^2} \leq 1 + \frac{1}{2h' - i}.
		\end{align*}
		The first inequality used the induction hypothesis and the second one used that for every $\theta \in [0, 1]$, we have $e^\theta \leq 1 + \theta + \theta^2$. 
	\end{proof}
	
	So, by Lemma~\ref{lemma:bound-exp-mp} and~\ref{lemma:bound-alpha}, we have $\E[\bfe^{sm_{\bfr}}] \leq \alpha_{h'}^1 \leq 1 + \frac{1}{h'} = 1 +O\left(\frac{1}{\log n}\right) $. 
	This finishes the proof of Property~(\ref{thm:main}c) in Theorem~\ref{thm:main}.

\section{Bicriteria-Approximation Algorithm for Degree-Bounded Group Steiner Tree on Trees}
\label{sec:dbgst-t}

In this section, we prove Theorem~\ref{thm:dbgst-t}, which is repeated here. 
\thmdbgstt*

We first set up some notations for the theorem. Recall that $T^\circ$ is the input tree, ${V^\circ}$ denotes the set of vertices of $T^\circ$, and $r$ denotes the root of $T^\circ$. For simplicity, we assume the costs are on the vertices instead of edges: Every vertex $u \in V^\circ$ has a cost $c_u \geq 0$. Notice that this does not change the problem.   We have $k$ groups indexed by $[k]$. For each group $t \in [k]$, we are given a set $O_t \subseteq V^\circ$ of leaves in $T^\circ$. W.l.o.g, we assume all $O_t$'s are disjoint.  Every vertex $v \in V$ is given a degree bound $D_v$.  The goal of the problem is then to output the smallest cost subtree $T$ of $T^\circ$ that satisfies the degree constraints and contains the root $r$ and one vertex from each $O_t$, $t \in [k]$.  Since now we only have one tree $T^\circ$, we use the following notations for children and descendants: For every vertex $u \in V^\circ$, let $\Lambda_u$ denote the set of children of $u$ in $T^\circ$, and $\Lambda^*_u$ to denote the set of descendants of $u$ in $T^\circ$ (including $u$ itself).

Now we describe the LP relaxation we use for our problem.  For every vertex $u \in T^\circ$, we use $x_u$ to indicate whether $u$ is chosen or not (in the correspondent integer program).  LP~\eqref{LP:DB-GST-T} is a valid LP relaxation for the DB-GST-T problem: 
\begin{equation}
  \text{min} \qquad \sum_{u \in {V^\circ}}c_ux_u \qquad \text{s.t.} \label{LP:DB-GST-T}
\end{equation} \vspace*{-25pt}

\noindent\begin{minipage}[t]{0.5\textwidth}
	\begin{alignat}{2}
		x_{v} &\leq x_u &\qquad &\forall u \in {V^\circ}, v \in \Lambda_u \label{eq:DB-GST-T-monotone} \\
	    \sum_{o \in O_t} x_o &= 1 &\qquad  & \forall t\in[k] \label{eq:DB-GST-T-connectivity} \\ 
	     \sum_{o \in O_t \cap \Lambda^*_u} x_o &\leq x_u &\qquad  & \forall t\in[k], \forall u \in {V^\circ} \label{eq:DB-GST-T-capacity}
	 \end{alignat}
\end{minipage}
\begin{minipage}[t]{0.5\textwidth}
	 \begin{alignat}{2}
	    \sum_{v \in \Lambda_u}x_v &\leq d_u \cdot x_u &\qquad  & \forall u\in {V^\circ} \label{eq:DB-GST-T-degree}\\
	    x_u &\in [0, 1] &\qquad  & \forall u \in {V^\circ} \label{eq:DB-GST-T-0-1}
	\end{alignat}
\end{minipage}\smallskip

In the correspondent integer program, the objective we try to minimize is $\sum_{u\in {V^\circ}}c_u x_u$, i.e, the total cost of all verticies we choose.  Constraint~\eqref{eq:DB-GST-T-monotone} says that if we choose a vertex $v$ then we must choose its parent $u$. Constraint~\eqref{eq:DB-GST-T-connectivity} requires for every group $t$, exactly one vertex in $O_t$ is added to the tree. Constraint~\eqref{eq:DB-GST-T-capacity} holds since if $u$ is chosen, at most one vertex in $\Lambda^*_u \cap O_t$ is chosen for every group $t$. Constraint~\eqref{eq:DB-GST-T-degree} is the degree constraint.  In the LP relaxation, we require each $x_u$ to take value in $[0, 1]$ (Constraint~\eqref{eq:DB-GST-T-0-1}).  Notice that \eqref{eq:DB-GST-T-connectivity} and \eqref{eq:DB-GST-T-capacity} for the root $r$ imply that $x_r = 1$. 

\paragraph{Modifying the LP solutions.}
Solving LP~\eqref{LP:DB-GST-T}, we can obtain the optimum LP solution $(x_u)_{u \in V^\circ}$. In our rounding algorithm, it would be convenient if every $x_u$ is a (non-positive) integer power of $2$ that is not too small. So, we shall modify the LP solution using the following operations, which may violate many of the LP constraints slightly.  For every $v\in V^\circ$ with $x_v < \frac1{2n}$, we change $x_v$ to $0$. This can only decrease the cost of the solution. It is easy to see that Constraints~\eqref{eq:DB-GST-T-monotone}, \eqref{eq:DB-GST-T-capacity} and \eqref{eq:DB-GST-T-degree} will not be violated.  Constraint~\eqref{eq:DB-GST-T-connectivity} may not hold any more, but we still have $\sum_{v \in O_t}x_v \geq 1 - n \times \frac{1}{2n} \geq \frac12$ for every $t \in [k]$. We can remove all vertices $v$ with $x_v = 0$ from the instance and thus assume $x_v \geq \frac{1}{2n}$ for every $v \in V^\circ$. Next, we increase each $x_v$ to the smallest (non-positive) integer power of $2$ that is greater than or equal to $x_v$.  This will violate many constraints in the LP by a factor of $2$. We list the properties that our new vector $(x_u)_{u \in V^\circ}$ has:
\begin{enumerate}[label = (P\arabic*), itemsep=0pt]
	\item For every $u \in {V^\circ}$, $x_u$ is an integer power of $2$ between $\frac{1}{2n}$ and $1$. \label{property:DB-GST-T-power-of-2}
	\item The $x$ values along any root-to-leaf path in $T^\circ$ is non-increasing. \label{property:DB-GST-T-monotone}
	\item $\sum_{o \in O_t}x_o \in [\frac12, 2]$ for every group $t \in [k]$.  \label{property:DB-GST-T-connectivity}
	\item $\sum_{o \in O_t \cap \Lambda^*_u} x_o \leq 2x_u$ for every $ t\in[k]$ and $ u \in {V^\circ}$. \label{property:DB-GST-T-capacity}
	\item $\sum_{v \in \Lambda_u}x_v \leq 2d_u x_u$ for every $u\in {V^\circ}$.  \label{property:DB-GST-T-degree}
	\item $\sum_{u \in {V^\circ}} c_ux_u \leq 2\cdot \opt$, where $\opt$ is the cost of the optimum integer solution. \label{property:DB-GST-T-cost}
\end{enumerate}

\subsection{The rounding algorithm}
We now describe our rounding algorithm.  We define two important global parameters: $L := \ceil{\log (2n)}$ and $\gamma := \floor{\log L}-2$. 
%
We say an edge $(u,v)$ with $v \in \Lambda_u$ has ``hop value'' 1 if $x_u < x_v$ and $0$ if $x_u = x_v$. For every vertex $u \in {V^\circ}$, we define $\ell_u$ to be the sum of hop values over all edges in the path from the root to $u$ in $T^\circ$.  Thus,  for every $u \in {V^\circ}$ and $v \in \Lambda_u$, we have $\ell_v - \ell_u \in \{0, 1\}$, and $\ell_v = \ell_u$ if and only if $x_v = x_u$.  By Properties~\ref{property:DB-GST-T-power-of-2}  and \ref{property:DB-GST-T-monotone}, we have that $\ell_v \in [0, L]$ for every $v \in V^\circ$.

Our rounding algorithm is applied on some scaled solution $x'$, which is defined as follows:
\begin{align*}
	x'_u = 2^{\min\set{\ell_u, \gamma}}x_u, \text{for every } u \in {V^\circ}. 
\end{align*}
As we mentioned in the introduction, this change will increase the probability of choosing $v$ conditioned on choosing $u$ by a factor of $2$, for some $u \in V^\circ, v \in \Lambda_u$ with $\ell_u < \ell_v \leq \gamma$.

We prove one important property for $x'$, which is necessary for us to run the recursive rounding algorithm. 
\begin{claim}
	\label{claim:DB-GST-T-x'-monotone}
	For every $u \in {V^\circ}$ and $v \in \Lambda_u$, we have $x'_v \leq x'_u$.  
\end{claim}
\begin{proof}
	If $x_v = x_u$ then we have $(u, v)$ has hop value $0$ and thus $\ell_v = \ell_u$. In this case we have $x'_v = x'_u$ as well.  Otherwise, we have $x_v \leq x_u/2$ and $h_v = h_u + 1$. So, $\min\set{h_v, \gamma} \leq \min\set{h_u, \gamma} + 1$ and therefore $x'_v \leq x'_u$.
\end{proof}

Notice that $x'_r = 1$ and every $x'_v$ is an integer power of $2$ between $2^{-L}$ and $1$. Our recursive rounding algorithm is run over $x'$. In the procedure recursive-rounding$(u)$, we add $u$ to our output  tree and do the following: for every $v \in \Lambda_u$, with probability $x'_v/x'_u$ independent of all other choices, we call recursive-rounding$(v)$.  In the root recursion, we shall call recursive-rounding$(r)$.

Our final algorithm will repeat the recursive procedure $M$ times independently, for a large enough $M = O(\log k)$.  Let $T_1, T_2, \cdots, T_M$ be the $M$ trees we obtained from the $M$ repetitions. Our final tree $T$ will be the union of the $M$ trees.

We first analyze the expected cost of $T$. First focus on the tree $T_1$. It is easy to see that the probability $u$ is chosen by $T_1$ is exactly $x'_u \leq 2^\gamma x_u = O(L) x_u$. Therefore, the expected cost of $T_1$ is at most $O(L) \cdot \opt$ by Property~\ref{property:DB-GST-T-cost}. Therefore, the expected cost of the tree $T$ is at most $O\left(ML\right)\cdot \opt = O(L \log k)\cdot \opt = O(\log n \log k) \cdot \opt$.

We then analyze the degree constraints on $T$. Given that $u$ is selected by $T_1$, the probability that we select a child of $v$ of $u$ is $\frac{x'_v}{x'_u} \leq \frac{2x_v}{x_u}$.  By Property~\ref{property:DB-GST-T-degree}, we have $\sum_{v \in \Lambda_u}\frac{x'_v}{x'_u} \leq \sum_{v \in \Lambda_u}\frac{2x_v}{x_u} \leq 4d_u$. Consider all the $M$ trees $T_1, T_2, \cdots, T_M$. Even if we condition on the event that $u$ appears in all the $M$ trees, the degree of $u$ is the summation of many independent random $\{0, 1\}$-variables.  The expectation of the summation is at most $4M d_u = O(\log k)\cdot d_u$. Using Chernoff bound, one can show that the probability that the degree of $u$ is more than $O(\log n) \cdot d_u$ is at most $\frac{1}{10n}$, for some large enough $O(\log n)$ factor.  Therefore, with probability at least $0.9$, every node $u$ in $T$ has degree at most $O(\log n) \cdot d_u$. Therefore, we proved that the degree violation factor of our algorithm is $O(\log n)$, as claimed in Theorem~\ref{thm:dbgst-t}.

\subsection{Analysis of connectivity probability}
It remains to show that with high probability, the tree $T$ contains a vertex from every group.  This is the goal of this section.  
Till the end of the section, we focus on the tree $T_1$ and a fixed group $t$.  For every vertex $u \in {V^\circ}$, we define $\bfE_u$ to be the event that $u$ is chosen by $T_1$.  Our goal is to give a lower bound on $\Pr[\bigvee_{o \in O_t} \bfE_o]$, i.e, the probability that some vertex in $O_t$ is chosen by the tree $T_1$.

Notice that when two adjacent nodes in $T^\circ$ have the same $x'$ value, then the child is chosen whenever the parent  is. Thus, we can w.l.o.g contract any sub-tree of nodes in $T^\circ$ with the same $x'$ value into one single super-vertex, without changing the rounding algorithm.  Notice that if two adjacent vertices $u \in {V^\circ}, v \in \Lambda_u$ have $\ell_u = \ell_v$ then we have $x_u = x_v$ and thus $x'_u = x'_v$. So, we contract every maximal sub-tree of vertices in $T^\circ$ with the same $\ell$ value. After this operation, for every $u \in V^\circ$, $\ell_u$ is exactly the level of $u$ in the tree $T^\circ$. So, for every $u \in {V^\circ}$ and $v \in \Lambda_v$ we have $\ell_v = \ell_u + 1$. A super-vertex is in $O_t$ if one of its vertices before contracting is in $O_t$.  If an internal super-vertex is in $O_t$, we can remove all its descendants without changing the analysis in this section. So, again we have that $O_t$ only contains leaves.

For every vertex $u$, we define $$z_u = \sum_{o \in O_t \cap \Lambda^*_u}x_o.$$ Notice that $z_u \leq 2x_u$ by Property~\ref{property:DB-GST-T-capacity}.   

In the following, we shall bound $\Pr\Big[\bigvee_{o \in O_t \cap \Lambda^*_u}\bfE_o \big|\bfE_u\Big]$ for every $u \in V^\circ$ from bottom to top. This is done in two stages due to the threshold $\gamma$ we used when we define $x'$ variables. First we consider the case when $\ell_u \geq \gamma$ and then we focus on the case when $\ell_u < \gamma$. The two stages are captured by Lemmas~\ref{lemma:DB-GST-T-below-gamma} and \ref{lemma:DB-GST-T-above-gamma} respectively.
\begin{lemma}
	\label{lemma:DB-GST-T-below-gamma}
	For a vertex $u$ with $\ell_u \geq \gamma$, we have $\Pr\Big[\bigvee_{o \in O_t \cap \Lambda^*_u}\bfE_o \big|\bfE_u\Big] \geq \frac1{2(L + 1 - \ell_u)}\frac{z_u}{x_u}$. 
\end{lemma}
Similar lemmas have been proved multiple times in many previous results. Since our parameters are slightly different, we provide the complete proof here. There are two different approaches to prove the lemma, one based on bounding the conditional second moment of the random variable for the number of chosen vertices in $O_t \cap \Lambda^*_u$, and the other based on the mathematical induction on $\ell_u$, which is the one we use here.

\begin{proof}[Proof of Lemma \ref{lemma:DB-GST-T-below-gamma}]   Suppose $u$ is a leaf. Then $z_u/x_u = 1$ if $u \in O_t$ and $z_u/x_u = 0$ otherwise. So, we have $\Pr\Big[\bigvee_{o \in O_t \cap \Lambda^*_u}\bfE_o \big|\bfE_u\Big] = \frac{z_u}{x_u}$ and the lemma clearly holds since we have $\ell_u \leq L$. 

Then, we prove the lemma by induction on $\ell_u$. If $\ell_u = L$ then $u$ must be a leaf and thus the lemma holds. We assume the lemma holds for every $u$ with $\ell_u = \ell + 1$, for some $\ell \in [\gamma, L - 1]$.  Then we prove the lemma for $u$ with $\ell_u = \ell$.  If $u$ is a leaf the lemma holds and thus we assume $u$ is not a leaf. 
	\begin{align*}
		\Pr\Big[\bigvee_{o \in O_t \cap \Lambda^*_u}\bfE_o \big|\bfE_u\Big] &\geq 1-\prod_{v \in \Lambda_u}\left(1 - \frac{x'_v}{x'_u} \cdot \frac{1}{2(L-\ell)}  \frac{z_v}{x_v}\right) = 1-\prod_{v \in \Lambda_u}\left(1 - \frac{x_v}{x_u}\cdot \frac{1}{2(L-\ell)} \cdot \frac{z_v}{x_v}\right) \\
		&\geq 1 - \prod_{v \in \Lambda_u}\exp\left(-\frac{1}{2(L-\ell)} \cdot \frac{z_v}{x_u}\right) = 1 - \exp\left(-\frac{1}{2(L-\ell)}\cdot  \frac{z_u}{x_u}\right) \\
		&\geq \frac{1}{2(L-\ell)} \cdot \frac{z_u}{x_u} - \frac12\left(\frac{1}{2(L-\ell)} \cdot \frac{z_u}{x_u}\right)^2 \geq \frac{1}{2(L-\ell)}\cdot \frac{z_u}{x_u} - \left(\frac{1}{2(L-\ell)} \right)^2\frac{z_u}{x_u} \\
		&= \left(\frac{2(L-\ell)-1}{(2(L-\ell))^2}\right)\frac{z_u}{x_u} \geq \frac{1}{2(L+1-\ell)}\cdot \frac{z_u}{x_u}.
	\end{align*}
	The inequality in the first line used the induction hypothesis: $\frac{x'_v}{x'_u}$ is the probability that we choose $v$ in $T_1$ conditioned on that we choose $u$, and $\frac{1}{2(L-\ell)}  \frac{z_v}{x_v}$ is the lower bound on the probability that we choose some vertex in $O_t \cap \Lambda^*_v$ conditioned on that $v$ is chosen.  The equality in the line used that $x'_u = 2^\gamma x_u$ and $x'_v = 2^\gamma x_v$.  The inequality in the second line used that $1-\theta \leq e^{-\theta}$ for every real number $\theta$.  The first inequality in the third line used that $e^{-\theta} \leq 1-\theta + \frac {\theta^2}2$ for every $\theta \geq 0$.  The second inequality in the line used Property~\ref{property:DB-GST-T-capacity}, which says $\frac{z_u}{x_u}\leq 2$. The last inequality used that $(2(L-\ell)-1)\cdot 2(L-\ell +1) \geq 4(L-\ell)^2$ since $L -\ell \geq 1$.
\end{proof}

The lemma implies that for every $u$ with $\ell_u \geq \gamma$, we have $\Pr\Big[\bigvee_{o \in O_t \cap \Lambda^*_u}\bfE_o \big|\bfE_u\Big]\geq \frac{1}{2L}\cdot \frac{z_u}{x_u}$.  \smallskip

Now we analyze the probability for $u$ with $\ell_u \leq \gamma$. Recall that $\gamma = \floor{\log L } - 2$ and thus we have $2^\gamma \in (L/8, L/4]$.   Let $\alpha_\gamma = \frac{1}{2L}$ and for every $\ell \in [0, \gamma-1]$, define $\alpha_\ell = 2\alpha_{\ell+1} - 4\alpha^2_{\ell+1}$. It is easy to see that for every $\ell \in [0, \gamma]$, we have $\alpha_\ell \leq \frac{2^{\gamma - \ell}}{2L}$. Then, we have for every $\ell \in [0, \gamma-1]$, 
\begin{align*}
	\alpha_{\ell} = 2\alpha_{\ell+ 1} - 4\alpha_{\ell+1}^2  = 2\alpha_{\ell+1}(1-2\alpha_{\ell+1}) \geq 2\alpha_{\ell+1}\left(1 - 2\times  \frac{2^{\gamma-\ell - 1}}{2L}\right) = 2\alpha_{\ell+1}\left(1 -  \frac{2^{\gamma-\ell - 1}}{L}\right).
\end{align*}
Therefore, we have 
\begin{align*}
	\alpha_0 &\geq 2^\gamma \prod_{\ell=1}^{\gamma}\left(1-\frac{2^{\gamma-\ell-1}}{L}\right)\alpha_\gamma \geq \frac{2^\gamma}{2L}\prod_{\ell=1}^{\gamma}e^{-2^{\gamma-\ell}/L} \geq \frac{2^\gamma}{2L} e^{-2^{\gamma}/L} = \Omega(1).
\end{align*}
The second inequality used that $1-\theta \geq e^{-2\theta}$ for every $\theta \in  (0, 1/2)$. The last equality used that $\gamma = \floor{\log L} - 2$ and thus $2^{\gamma} = \Theta(L)$.

With the $\alpha$ values defined, we prove the following lemma via mathematical induction:
\begin{lemma}
		\label{lemma:DB-GST-T-above-gamma}
	For every vertex $\ell_u = \ell \leq \gamma$, we have $\Pr\Big[\bigvee_{o \in O_t \cap \Lambda^*_u}\bfE_o \big|\bfE_u\Big] \geq \alpha_{\ell}\frac{z_u}{x_u}$.
\end{lemma}
\begin{proof}
	The lemma holds if $\ell = \gamma$ as we mentioned. So, we assume $\ell < \gamma$ and the lemma holds with $\ell$ replaced by $\ell+1$.  If $u$ is a leaf, then we have $\Pr\Big[\bigvee_{o \in O_t \cap \Lambda^*_u}\bfE_o \big|\bfE_u\Big] = \frac{z_u}{x_u}$ and the lemma holds. So again we assume $u$ is not a leaf. Then, 
	\begin{align*}
		\Pr\Big[\bigvee_{o \in O_t \cap \Lambda^*_u}\bfE_o \big|\bfE_u\Big] &\geq 1-\prod_{v \in \Lambda_u}\left(1 - \frac{x'_v}{x'_u} \alpha_{\ell+1} \frac{z_v}{x_v}\right) = 1-\prod_{v \in \Lambda_u}\left(1 - \frac{2x_v}{x_u} \alpha_{\ell+1} \frac{z_v}{x_v}\right) \\
		&\geq 1 - \prod_{v \in \Lambda_u}\exp\left(-2\alpha_{\ell+1} \frac{z_v}{x_u}\right) = 1 - \exp\left(-2\alpha_{\ell+1}\frac{z_u}{x_u}\right) \\
		&\geq 2\alpha_{\ell+1}\frac{z_u}{x_u} - \frac12\left(2\alpha_{\ell+1}\frac{z_u}{x_u}\right)^2 \geq 2\alpha_{\ell+1}\frac{z_u}{x_u} - (2\alpha_{\ell+1})^2\frac{z_u}{x_u} = \alpha_{\ell}\frac{z_u}{x_u}.
	\end{align*}
	To see the equality in the first line, we notice that $x'_u = 2^{\ell}x_u$ and $x'_v = 2^{\ell+1}x_v$ for every $v \in \Lambda_u$.  Many other inequalities used the same arguments as in Lemma~\ref{lemma:DB-GST-T-below-gamma}.
\end{proof}

Applying the lemma for the root  $r$ of $T^\circ$, we have that $\Pr\big[\bigvee_{o \in O_t}\bfE_o\big] \geq \alpha_0 \cdot \frac{z_r}{x_r} \geq \alpha_0 \cdot \frac12 = \Omega(1)$.

Now we consider all the $M$ trees $T_1, T_2, \cdots, T_M$ together. The probability that $O_t$ is not chosen by any of the $M$ trees is at most $ \left(1 - \Omega(1)\right)^M \leq \frac{1}{10k}$, if our $M = O(\log k)$ is big enough.  Thus the probability that $T$, the union of all trees $T_1, T_2, \cdots, T_M$, contains an $r$-to-$O_t$ path for every $t$, is at least $0.9$.

\paragraph*{Acknowledgement}
X. Guo, S. Li and J. Xian are partially supported by NSF grants CCF-1566356, CCF- 1717134, CCF-1844890.
B. Laekhanukit is partially supported by Science and Technology Innovation 2030 –“New Generation of Artificial Intelligence” Major Project No.(2018AAA0100903), NSFC grant 61932002, Program for Innovative Research Team of Shanghai University of Finance and Economics (IRTSHUFE) and the Fundamental Research Funds for the Central Universities and by the 1000-talent award by the Chinese Government.
Daniel Vaz has been supported by the Alexander von Humboldt Foundation with funds from the German Federal Ministry of Education and Research (BMBF).

\bibliographystyle{plain}
\bibliography{dst}

\appendix
\section{Omitted Proofs}
\label{sec:appendix}
\begin{proof}[Proof of Lemma~\ref{lemma:balanced-partition}]
	We assume $n \geq 4$; otherwise, if $n = 3$, then we have $2n/3 + 1 = 3$, and $\root(T)$ satisfies the condition.   Our goal is to find a vertex $u$ with $n/3 < |\Lambda^*(u)| \leq 2n/3+1$.  Start from  $u = \root(T)$ in the tree, and thus, we have $\Lambda^*(u) > 2n/3 + 1$. Let $v$ be the child of $u$ with the biggest $|\Lambda^*(v)|$. So, $|\Lambda^*(v)| \geq (|\Lambda^*(u)|-1)/2> n/3$. We then replace $u$ with $v$. So $|\lambda^*(u)|$ has decreased but the condition $|\Lambda^*(u)| > n/3$ is maintained. Thus, if we repeat the process, we will eventually find a $u$ with $n/3< |\Lambda^*(u)| \leq 2n/3+ 1$.
	%
\end{proof}

\end{document}